\documentclass[aip,jcp,reprint,amsmath,amssymb,floatfix,citeautoscript]{revtex4-1}
\usepackage{amsmath}
\usepackage{graphicx}
\usepackage{amssymb}
\usepackage{amsthm}
\usepackage{bm}

\usepackage{ragged2e}
\usepackage[labelsep=period,format=plain,justification=justified,font=footnotesize]{caption}
\DeclareCaptionJustification{myjust}{\justifying}
\usepackage{times}
\usepackage{txfonts}

\newcommand{\vr}{{\mathbf{r}}}

\newcommand{\vQ}{{\mathbf{Q}}}

\begin{document}

\title{Microscopic theory of singlet exciton fission. I. General formulation
}

\author{Timothy C. Berkelbach}
\email{tcb2112@columbia.edu}
\affiliation{Department of Chemistry, Columbia University, 3000 Broadway, New York, New York 10027, USA}

\author{Mark S. Hybertsen}
\email{mhyberts@bnl.gov}
\affiliation{Center for Functional Nanomaterials, Brookhaven National Laboratory, Upton, New York 11973-5000, USA}

\author{David R. Reichman}
\email{drr2103@columbia.edu}
\affiliation{Department of Chemistry, Columbia University, 3000 Broadway, New York, New York 10027, USA}

\begin{abstract}

Singlet fission, a spin-allowed energy transfer process generating two triplet excitons from one singlet exciton,
has the potential to dramatically increase the efficiency of organic solar cells.  However, the dynamical mechanism
of this phenomenon is not fully understood and a complete, microscopic theory of singlet fission is lacking.  In this work,
we assemble the components of a comprehensive microscopic theory of singlet fission that connects excited state quantum chemistry
calculations with finite-temperature quantum relaxation theory.  We elaborate on the distinction between localized
diabatic and delocalized adiabatic bases for the interpretation of singlet fission experiments in both the
time and frequency domains.  We discuss various approximations to the exact density matrix dynamics and propose
Redfield theory as an ideal compromise between speed and accuracy for the detailed investigation of
singlet fission in dimers, clusters, and crystals.  Investigations of small model systems based on parameters typical of
singlet fission demonstrate
the numerical accuracy and practical utility of this approach.

\end{abstract}

\maketitle

\section{Introduction}
The Shockley-Queisser limit places the maximal efficiency of a single-junction
solar cell at about 31\%\cite{sho61}.  Promising technologies aimed at exceeding this limit
include tandem solar cells\cite{dev80,kim07}, hot carrier collection\cite{ros82,ros93,noz01}, and
multiple exciton generation (MEG)\cite{wer94,sem11}.  In MEG and its molecular analogue, singlet fission, a single absorbed photon
generates two or more excitons each of lower energy, eventually yielding two or more electron-hole pairs.  
This mechanism results in theoretical solar cell efficiencies of almost 50\% with singlet fission\cite{smi10} or
more with MEG\cite{wer94}. Singlet fission
is a particularly promising technology in inexpensive organic solar cells, whose efficiencies to date
remain well below that of their more expensive inorganic counterparts.  Proposals to utilize singlet fission in this manner
have targeted covalently linked dimers for use in dye-sensitized cells\cite{pac06} as well as crystalline
materials for more traditional heterojunction cells\cite{jad11}.

Despite initial reports of singlet fission over 40 years ago\cite{sin65,swe68,gea69,mer69}, an explosion of experimental studies have emerged only
recently due to the aforementioned potential for photovoltaic utility.  Singlet fission, as typically measured by the observation
of triplets in the form of delayed fluorescence (DF) or transient absorption (TA), has been found to vary from system to system 
both in total yield and overall timescale, making the search for unifying principles very difficult.  The authoritative
review by Smith and Michl\cite{smi10} effectively summarizes the state of the field up to 2010.  Since then, singlet
fission has been further investigated by TA in thin films of diphenylisobenzofuran\cite{joh10}, by DF and TA in 
crystalline tetracene\cite{bur11,bur12}, by time-resolved two-photon photoemission\cite{cha11} and TA\cite{wil11}
in crystalline pentacene, by TA in solution and crystalline rubrene\cite{ma12},
and even by DF and TA in amorphous films of diphenyl tetracene\cite{rob12}.  Although
still far away from commercial use in solar cells, singlet fission has been investigated 
in pentacene-perylene blend films\cite{ram12}
and even successfully incorporated into heterojunction
solar cells utilizing phthalocyanine, tetracene, and C$_{60}\cite{jad11}$.

These many enlightening experiments notwithstanding, the dynamical
mechanism of singlet fission is still not well understood.  Previous theoretical work has focused almost entirely on identifying
the quantum mechanical states involved, including their wavefunction character and energetic ordering.  
Perhaps most significantly, high-level quantum chemistry calculations have identified a multi-exciton
state that is composed of two triplets coupled into an overall spin singlet\cite{zim10,zim11}. The transition
to this multi-exciton state is thus spin-allowed and should proceed rapidly, while its triplet-triplet character
suggests that it should naturally relax to separated triplets on a longer timescale. Unfortunately, the multi-exciton
nature of this state implicitly prevents its direct photoexcitation such that the state is spectroscopically ``dark'' and difficult to
observe. However, time-resolved two-photon photoemission\cite{cha11} 
and transient absorption\cite{bur12} spectroscopic measurements have
provided direct and indirect evidence, respectively, of this multi-exciton state in ultrafast singlet fission.

A simple four-electron four-orbital model suggests at least two viable mechanisms for the transition from an initially excited intramolecular
singlet state, $S_1$, to the multi-exciton triplet-triplet state, $TT$\cite{smi10}.
The first is a \textit{mediated} mechanism, whereby a
charge transfer state acts as an intermediate in the transition from $S_1$ to $TT$; theoretical studies of this mechanism in
coupled molecular dimers have considered
the static electronic parameters\cite{gre10_jpcb2}, as well as the real-time dynamics in the limit of fast coherent
transfer\cite{gre10_jpcb1} and in the presence of a low-frequency
solvent bath\cite{tei12}.  Alternatively, a \textit{direct} mechanism has also been implicated, whereby the Coulomb potential
yields a direct interaction between $S_1$ and $TT$, avoiding any intermediates.  Some authors have invoked such a proposal
to explain fission in crystalline tetracene and pentacene, based both on experiment\cite{cha11} and quantum chemistry calculations
of clusters\cite{zim11}.

An internally consistent theory of singlet fission phenomena must comprise a currently nonexistent unification of static
electronic structure and dynamic relaxation mechanisms.
In this first article of a series we pursue this goal,
presenting a fully microscopic theoretical formalism tailored to the investigation of
singlet fission in molecular systems.
Our goal is the identification, extension,
and marriage of existing techniques of electronic structure theory and microscopic quantum dynamics for the efficient
and accurate treatment of singlet fission in organic molecules and bulk materials.  Such a synthesis elucidates experimental results
in both the time and frequency domains and allows for studies of competing mechanisms as well as quantitative predictions.
Our approach is related in spirit to treatments of excitation energy transfer
in photosynthetic pigment protein complexes, where reduced density matrix simulations similar to those proposed
here have enjoyed great success in understanding quantum effects in complex, multi-state biological systems\cite{ish09_jcp2,ish09_pnas,reb09}.
The second and third articles of this series will make clear the utility of this formalism as we investigate singlet fission in
dimers and crystals, respectively.

The layout of this paper is as follows.  In Sec.~\ref{sec:elec}, we present a minimal electronic structure model capable
of describing all relevant states and couplings for the problem of singlet fission.  We then proceed in Sec.~\ref{sec:bath}
to describe a non-Markovian quantum master equation approach for the description of relaxation mechanisms arising from the
coupling of electronic degrees of freedom with nuclear vibrations in systems undergoing singlet fission.
We present numerical examples of our approach in Sec.~\ref{sec:redfield} as applied to simple model systems of singlet
fission, benchmarking our results against numerically exact calculations.
In Sec.~\ref{sec:conc}, we reflect on our approach and conclude.

\section{Electronic structure of singlet fission}\label{sec:elec}

Our theoretical framework begins with the electronic structure of singlet fission chromophore systems in the
limit of frozen nuclei.  In Sec.~\ref{ssec:basis}, we will emphasize the utility of a generically defined
diabatic basis of excited states, and in particular how they should be interpreted, thereby providing a rigorous and important language
with which to speak about the electronic structure of singlet fission.
As a practical, physically intuitive \textit{example} of such a framework, in Sec.~\ref{ssec:minimal}, we will present a limited configuration interaction (CI)
description of these diabatic excited states.
This model quantum chemical formalism will be recognized as a modest generalization of the picture 
proposed by Smith and Michl\cite{smi10} to understand singlet fission in dimers.  Our contributions
are to distinguish between the diabatic basis
and the adiabatic basis, as related to both theoretical development and experimental interpretation,
to formalize this procedure within the context of {\em ab initio} quantum chemistry,
and to generalize to the case of more than two molecules.  We will briefly discuss alternative quantum chemistry approaches in Sec.~\ref{ssec:aside}
before assessing the accuracy and summarizing the proposed electronic structure formalism in Sec.~\ref{ssec:acc}.

\subsection{The basis of diabatic states}\label{ssec:basis}

Consider $M$ molecules with a total of $N$ electrons.
We start by defining a basis of \textit{diabatic} electronic states, i.e. those states whose quantum mechanical character is well-defined
and presumed to be independent of the molecular geometry.  For an accessible discussion of diabatic states in the context of electron transfer,
see Ref.~\onlinecite{van10}.  

The desired basis begins with the exact many-electron wavefunction corresponding to the (singlet) ground state of the system,
$\Psi_{S_0}(\vr_1,\vr_2,\dots,\vr_N) \equiv \langle \vr | S_0 \rangle$.  All remaining states in the minimal singlet fission diabatic basis
are excitations above the ground state, although they are not in general eigenstates of the electronic Hamiltonian.
Specifically, the basis must include all $M$ excited states which may be
characterized by molecule $m$ being in its first excited singlet (or $S_1$) state, $|(S_1)_m\rangle$, sometimes referred to as Frenkel excitations.
The next class of excited states are charge-transfer (CT) states,
where molecule $m$ has a single positive charge and molecule $n$ has a single negative charge, denoted $|C_m A_n\rangle$
($C$ for cation and $A$ for anion). The final class of necessary excited states are those multi-exciton states described as being a spin-adapted
combination of triplet excitations on molecules $m$ and $n$, forming an overall singlet, $|T_m T_n\rangle$.

Double excitations that instead couple two singlets are presumed too high in energy to be relevant for singlet
fission and are consequently neglected, as are all double excitations involving more than two molecules.
States of differing multiplicity (such as triplet- and quintet-coupled triplets, $^3TT$ and $^5TT$) can also
be included.  While a purely electronic Hamiltonian does not couple
such states of different multiplicity, the spin dipole-dipole Hamiltonian does\cite{smi10}. We neglect this Hamiltonian in
our present formalism because it is significantly weaker, and we instead only focus on the short-time formation of
the spin-singlet multi-exciton state, $|T_mT_n\rangle$.  However, including spin dipole-dipole and Zeeman interactions in the
presence of a magnetic field would be important but straightforward modifications to study such long-time effects.

Although these many-electron basis states, which we will denote generically by $|i\rangle$ and $|j\rangle$,
are not eigenstates of the electronic Hamiltonian and do not constitute a complete basis, one may consider the projection
of the true electronic Hamiltonian onto this basis,
\begin{equation}
\hat{H}_{el} \approx \sum_{ij} |i\rangle \langle i | \hat{H}_{el} | j \rangle \langle j |.
\end{equation}
We wish to emphasize that the states defined above merely constitute a \textit{physically-motivated
(diabatic) basis}.  Many experimental measurements instead observe the \textit{adiabatic} basis, which is 
is approximately obtained from
the diagonalization of the above electronic Hamiltonian, $\hat{H}_{el}$, yielding eigenstates that are a mixture
of the diabatic states.\footnote{More accurately, the eigenstates should be
those of the total Hamiltonian, which includes coupling to nuclear motion.}  Therefore, one must
exercise great caution when speaking about the character of observed states, e.g. ``charge-transfer,'' and when
discussing results and proposing mechanisms in terms of these states.  This distinction must also be kept in mind for
traditional excited state electronic structure calculations, which inherently probe the adiabatic, and not the diabatic basis.

In light of the above proviso, one may naturally question the utility of this diabatic basis.  We propose three
reasons to begin the theoretical development of singlet fission from this basis:

i.) The physical character of the basis
states aids in \textit{interpreting} the nature of observed eigenstates, allowing for a means to quantify statements such as ``a mixture of
charge-transfer and Frenkel excitations.'' This latter example will play a prominent role in our future work on
singlet fission in crystals.  Similarly, this principle underlies the coherent superposition approximation recently
proposed to explain MEG in nanocrystals\cite{ell05,sha06} and singlet fission in pentacene\cite{cha11}, wherein
single- and multi-exciton (diabatic) states are coupled to yield an eigenstate that is a superposition of the two.

ii.) The local diabatic basis can yield accurate results which computationally scale very favorably.
Proximity arguments alone can naturally suggest coupling terms that may be approximated or neglected entirely.
Using such
approximations, to be discussed in more detail in our next paper,
one may easily build up a large molecular aggregate Hamiltonian
using only
diabatic energies and couplings from monomers, dimers, or small clusters, which may be computed with very high accuracy. This
philosophy is reminiscent of fragmentation methods in the pursuit of linear scaling quantum chemistry\cite{gor12}.

iii.) Lastly, the molecular character of the diabatic basis allows for a straightforward extension to include coupling to molecular
vibrations, which naturally separate into intramolecular and intermolecular modes,  as we detail in the following section.

Clearly, the accurate construction of diabatic states marks an important research goal for
\textit{ab initio} simulations of singlet fission.  While our approach here and henceforth employs a constructive strategy,
i.e. a direct construction of diabats without explicit reference to the adiabatic states of the extended system,
an alternative route would employ deductive strategies that 
attempt to obtain approximate diabats given a set of adiabats. This latter set of states is more easily
obtained at high accuracy from existing quantum chemistry methods, although the non-uniqueness of this
diabatization procedure results in various competing methods with subtle differences\cite{fos60,edm63,van10}.
In any case, the framework presented here is not limited to the CI-type model Hamiltonian outlined below,
and more accurate diabatic states, as might be obtained from multi-reference quantum chemistry methods,
can be naturally incorporated into the
dynamical scheme to be discussed in Sec.~\ref{sec:bath}.

\subsection{A minimal, truncated CI basis}\label{ssec:minimal}

The accurate quantum mechanical calculation of
excited states in large molecular systems is still a difficult challenge (see Refs.~\onlinecite{zim10,zim11} for examples of recent high-level quantum
chemistry calculations as applied to singlet fission) and thus we consider here the simplest possible model
Hamiltonian approach that captures the essential physics contained in the diabatic framework outlined above.
Specifically, we consider the minimal active space of all Hartree-Fock (HF), or HF-like,
highest occupied and lowest unoccupied molecular orbitals (HOMOs and LUMOs) of the \textit{isolated} molecules; extension to include
additional frontier orbitals is straightforward.  We
furthermore restrict the electronic structure calculation to all single and select double excitations, the latter ensuring treatment of the
bi-excitonic triplet-triplet state.  If done as a purely \textit{ab initio} theory, this approach would be somewhat akin to
configuration interaction\cite{sza96} with single and (select) double excitations (CISD) with the frozen core and deleted virtuals approximations,
or alternatively a type of (severely) restricted active space CISD.  However, our formalism differs slightly in that we consider excitations among the isolated
molecular orbitals, rather than among the HF orbitals of the full interacting system.

To make our description more precise, we define the creation (annihilation) operator for the HOMO of molecule $m$
with spin $\sigma$ as
$c^\dagger_{H,m,\sigma}$ ($c_{H,m,\sigma}$) and likewise for the LUMO.  The ground state is thus taken to be
\begin{equation}
|S_0\rangle = \prod_{m=1}^{M}\prod_{\sigma=\uparrow,\downarrow} c^\dagger_{H,m,\sigma} |0\rangle
\end{equation}
where $|0\rangle$ is the vacuum state of inactive core orbitals, thus filling the HOMO of all molecules. As discussed above, 
this state is not the result of a self-consistent HF procedure.
From this ground state, we will generate the three types of excited states described above in Sec.~\ref{ssec:basis}.  
Because the electronic Hamiltonian is spin-conserving, we take symmetry-adapted linear combinations of select excitations
to generate simultaneous eigenstates of the $\hat{S}_z$ and $\hat{S}^2$ operators, with
eigenvalues of 0 for both (sometimes called configuration state functions).

The first type of state is the local Frenkel singlet excitation on molecule $m$, given by
\begin{equation}
|(S_1)_m\rangle = \frac{1}{\sqrt{2}}\left( c^\dagger_{L,m,\uparrow}c_{H,m,\uparrow}
	+ c^\dagger_{L,m,\downarrow}c_{H,m,\downarrow} \right) |S_0\rangle.
\end{equation}
In addition to the above intramolecular excitation, the single excitations also generate our second type of
state, namely the intermolecular charge-transfer excitation obtained by exciting an electron from the
HOMO of molecule $m$ to the LUMO of molecule $n$
($n\neq m$),
\begin{equation}
|C_m A_n\rangle = \frac{1}{\sqrt{2}}\left( c^\dagger_{L,n,\uparrow}c_{H,m,\uparrow}
	+ c^\dagger_{L,n,\downarrow}c_{H,m,\downarrow} \right) |S_0\rangle,
\end{equation}
where $C$ and $A$ denote the cationic and anionic species, respectively.
The above two types of excited states combine to yield all possible single excitations, so that stopping
at this point would constitute a full CI-singles (CIS) within the HOMO-LUMO space.

However as discussed above, the problem of singlet fission necessarily requires our third type of state, a double excitation
coupling two intramolecular triplet excitations into a state with overall singlet character,
\begin{widetext}
\begin{equation}
\begin{split}
|T_m T_n\rangle &= \frac{1}{\sqrt{12}}
	\Big[
	2c^\dagger_{L,n,\downarrow} c^\dagger_{L,m,\uparrow} c_{H,n,\uparrow} c_{H,m,\downarrow}
	+ 2c^\dagger_{L,n,\uparrow} c^\dagger_{L,m,\downarrow} c_{H,n,\downarrow} c_{H,m,\uparrow}
	- c^\dagger_{L,m,\downarrow} c^\dagger_{L,n,\downarrow} c_{H,n,\downarrow} c_{H,m,\downarrow} \\
	&\hspace{4em} + c^\dagger_{L,m,\downarrow} c^\dagger_{L,n,\uparrow} c_{H,n,\uparrow} c_{H,m,\downarrow}
	+ c^\dagger_{L,m,\uparrow} c^\dagger_{L,n,\downarrow} c_{H,n,\downarrow} c_{H,m,\uparrow}
	- c^\dagger_{L,m,\uparrow} c^\dagger_{L,n,\uparrow} c_{H,n,\uparrow} c_{H,m,\uparrow}
	\Big] |S_0\rangle.
\end{split}
\end{equation}
\end{widetext}

Finally, we point out that because the molecular orbitals of distinct isolated molecules are not necessarily orthogonal
to one another, the use of creation and annihilation operators acting in the space of these orbitals is not strictly rigorous.
While one could imagine employing suitably orthogonalized molecular orbitals that retain the localized nature of isolated
orbitals, the actual overlap in molecular dimers and crystals is often negligibly small, thus justifying the theory in
its present form.

Having defined a set of diabatic basis states, it thus remains to calculate all matrix elements of the electronic Hamiltonian,
$\langle i | \hat{H}_{el} | j \rangle$.
While the calculation is straightforward, the results are cumbersome, and so we include the explicit results in
App.~\ref{app:basis}. As discussed more below in Sec.~\ref{ssec:acc}, the diagonal matrix elements (energies) are only expected to be of
qualitative accuracy but can provide useful insight, and likewise for the off-diagonal elements (couplings).  For example,
the couplings naturally separate into two classes: those containing one-electron integrals and
those containing only two-electron integrals.  The one-electron integrals include the simple kinetic
energy term, describing favorable charge delocalization, or ``hopping.''  Such one-electron integrals
are expected to be one or more orders of magnitude larger than the two-electron ones.  Reasonable
estimates for typical singlet fission chromophores in close proximity
are 50-100 meV for one-electron integrals and 5 meV or less
for two-electron integrals.  These simple analytical expressions and order of magnitude estimates
contribute to the interpretation of singlet fission in terms of mediated and direct mechanisms.
Qualitatively, the mediated mechanism proceeds via two one-electron processes, whereas the direct
mechanism proceeds by one two-electron process.  Which of these two diametric mechanisms prevails
in a given system of interest will depend sensitively on the relative energies of the diabatic states and the
dynamics of the nuclear degrees of freedom, as will be discussed in Sec.~\ref{sec:redfield}.

\subsection{Aside regarding wavefunction free methods}\label{ssec:aside}

Although the formalism here has employed the HF orbitals to construct a many-electron basis,
we pause to consider some alternatives.  At least at the level of single excitations, many other electronic
structure theories can be reduced to an eigenvalue equation for the transition energies, much like CIS.
Note that the CIS theory amounts to the diagonalization of an effective \textit{two-particle} Hamiltonian,
\begin{equation}\label{eq:2part}
H_{ij,kl} = \delta_{ik}\delta_{jl}\left( \varepsilon_j - \varepsilon_i \right)
    + \left(f_i - f_j\right) \mathcal{K}_{ij,kl},
\end{equation}
where $f_i$ is the ground-state occupancy of orbital $i$ and
\begin{equation}
\mathcal{K}_{ij,kl} = 2 (il|jk) - (il|\hat{W}(\vr_1,\vr_2)|kj).
\end{equation}
Clearly, for CIS, $\hat{W}(\vr_1,\vr_2) = r_{12}^{-1}.$
Physically, the vertex $\mathcal{K}$ describes the interaction between
single-particle excitations $i\rightarrow j$ and $k\rightarrow l$.
If the original single-particle
states are not a good approximation to the quasiparticles of the system, as determined e.g. by
comparison with electron affinity and ionization energies, then the HF excitations are in some
sense a poor starting point on which to build interactions.  In other words, the true many-body
excitations will require contributions from many single-particle excitations.

Instead, one could start from a ground state density functional theory (DFT) calculation of the isolated
molecules, and then consider excitations within the Kohn-Sham (KS) orbitals; we will not dwell here on the
physical reasons for which the KS orbitals may be better single particle states. 
Suffice it to say that this approach is adopted in time-dependent DFT (TD-DFT)\cite{hir99,mar04,cas12} and many-body Green's function
approaches\cite{hyb86,roh00,oni02}, both of which typically yield results superior to those of HF-based CIS, finding many-body
excitations strongly dominated by far fewer single-particle excitations.  For example,
the Green's function based Bethe-Salpeter equation (BSE) in practice adopts the form Eq.~(\ref{eq:2part}) with perturbatively corrected orbital energies and
a statically screened interaction,
\begin{equation}
\hat{W}(\vr_1,\vr_2) \approx \int d\vr \epsilon^{-1}(\vr_1,\vr,\omega=0) |\vr-\vr_2|^{-1},
\end{equation}
where $\epsilon(\vr_1,\vr_2,\omega)$ is the frequency-dependent dielectric function\cite{roh00}.  In the crude limit where
$\epsilon^{-1}(\vr_1,\vr_2,\omega=0) = \epsilon^{-1} \delta(\vr_1-\vr_2)$, with $\epsilon$ a dielectric constant,
one arrives at simply static screening of the direct Coulomb term,
\begin{equation}\label{eq:bbse}
\mathcal{K}_{ij,kl} = 2 (il|jk) - \epsilon^{-1} (il|kj).
\end{equation}
On the other hand, the inclusion
of doubly excited states presents an ongoing challenge to these former methodologies, making them difficult to employ
in an internally consistent theory of singlet fission.  Recent work has demonstrated that higher excitations only arise in the above theories
with the retention of a frequency dependent interaction kernel\cite{rom09,san11}. Specifically, one solves the eigenvalue-like equation,
$H(\omega) c = \omega c$, with
\begin{equation}
\begin{split}
H_{ij,kl}(\omega) &= \delta_{ik}\delta_{jl}\left(\epsilon_j - \epsilon_i\right) \\
&\hspace{1em} + \left(f_i - f_j\right)\left[ 2 (il|jk) - (il|\hat{W}(\vr_1,\vr_2,\omega)|kj) \right].
\end{split}
\end{equation}
The operator $\hat{W}(\omega)$ is related to the dynamically screened dielectric function in BSE and to the exchange-correlation kernel
in TD-DFT (i.e. the adiabatic approximation precludes observation of multiple excitations in TD-DFT).
We consider this a very interesting research focus for
singlet fission and a subject of future work.

\subsection{Accuracy and summary}\label{ssec:acc}

Based on the preceding discussion, the numerical results of the approach proposed in Sec.~\ref{ssec:minimal} will
only be of qualitative accuracy.  The diagonal energies, $\langle i | \hat{H}_{el} | i \rangle$, should be considered estimates of
their true values, with some leeway for semi-empirical adjustment.  For example, we find that the gas-phase $S_0 \rightarrow S_1$
transition energy of pentacene predicted by the above approach with a 6-31G(d) basis set is 3.76 eV, to be compared to
the experimental value of approximately 2.3 eV\cite{bie80,hei98}.  Including the additional dynamical correlation
arising from the frozen orbitals (i.e. not just the HOMO and LUMO) yields the improved value of 2.81 eV.
While TD-DFT is typically expected to be an improvement, it was shown previously to predict values of 1.64 eV and 1.90 eV, for the
PBE and B3LYP functionals, respectively\cite{kad06}.  Thus, even purportedly sophisticated methods yield excitation
energies with errors ranging from 0.4 to 0.7 eV\cite{gri03,kad06}.
Interestingly, the ad-hoc BSE-like prescription, Eq.~(\ref{eq:bbse}), using only the DFT HOMO and LUMO from B3LYP and
the dielectric constant of pentacene $\epsilon = 3.6$, predicts
a transition energy of 2.95 eV, much improved from the HOMO-LUMO CIS result (note that there is, however, no a priori reason
that the dielectric constant for bulk pentacene should be physically meaningful for a single molecule).
Only multi-reference perturbation theory\cite{zim10} and full many-body $GW$/BSE calculations\cite{sha12} yield quantitative
accuracy, predicting 2.1 and 2.2 eV, respectively.
Similarly, the electronic couplings in this basis may not be quantitatively accurate, but have already been
shown in other work to provide useful qualitative insight into the efficiency of singlet fission through investigation of their
\textit{relative} magnitudes\cite{gre10_jpcb2} and dependence on molecular orientation\cite{smi10}.  It may thus be
permissible to uniformly scale the electronic coupling matrix elements when investigating singlet fission.

To summarize, we argue that the diabatic basis, comprising states that are easily characterized and
energies and couplings that are straightforwardly calculated, acts as the crucial conceptual intermediate between
high-level quantum chemistry calculations, which inherently yield electronically adiabatic states that are difficult
to characterize, and microscopic quantum master equations, which are required to accurately treat thermally induced
relaxation effects, the topic of the next section.

\section{System-bath quantum dynamics}\label{sec:bath}

In this section, we consider the coupling of electronic (system) and nuclear (bath) degrees of freedom.  Although the treatment
is relatively standard and can be found in textbooks, see e.g. Ref.~\onlinecite{may11}, we include the derivation
in App.~\ref{app:sysbath},
to emphasize the microscopic connection to the diabatic basis introduced above. 
The result is the system-bath Hamiltonian described in Sec.~\ref{ssec:sbham}.

\subsection{System-bath Hamiltonian}\label{ssec:sbham}

To include the effects of electron-phonon coupling, we employ a system-bath type Hamiltonian
\begin{equation}\label{eq:ham_sb}
\hat{H}_{tot} = \hat{H}_{el} + \hat{H}_{el-ph} + \hat{H}_{ph}
\end{equation}
with the electronic Hamiltonian calculated at the ground-state geometry in terms of
the diabatic states described in Sec.~\ref{sec:elec},
\begin{equation}
\hat{H}_{el} = \sum_i |i\rangle E_i \langle i| + \sum_{ij} |i\rangle V_{ij} \langle j |,
\end{equation}
the bilinear electron-phonon coupling,
\begin{equation}
\hat{H}_{el-ph} = \sum_i |i\rangle\langle i | \sum_{k} c_{k,i} \hat{q}_k
	+ \sum_{ij} |i\rangle \langle j | \sum_{k} c_{k,ij} \hat{q}_k
\end{equation}
with $c_{k,0} = 0$, and the free phonon Hamiltonian,
\begin{equation}
\hat{H}_{ph} = \sum_{k} \left[ \frac{\hat{p}_k^2}{2} + \frac{1}{2}\omega_k^2 \hat{q}_k^2 \right].
\end{equation}
In the above, $i$ and $j$ index the diabatic electronic basis states, and $k$ indexes both the inter- and
intra-molecular (ground state) normal modes of the system.

The molecular vibrations and phonons are completely described by their spectral density,
\begin{equation}
J_{ij}(\omega) = \frac{\pi}{2} \sum_{k} \frac{c_{k,ij}^2}{\omega_k} \delta(\omega-\omega_k)
\end{equation}
Physically, the spectral density encodes the distribution of normal mode frequencies weighted by the strength with which
each mode couples to the energy level of diabatic state $i$ ($J_{ii}(\omega)$) or to the electronic coupling
between states $i$ and $j$ ($J_{ij}(\omega)$).
In practice, spectral densities (obtained in a manner to be described) are usually fit to a numerically convenient functional form,
$J(\omega) = \lambda F(\omega/\Omega)$, parametrized by the reorganization energy,
$\lambda = \pi^{-1}\int d\omega J(\omega)/\omega$ and a characteristic frequency $\Omega$.

Atomistically, the spectral densities may be calculated through a combination of classical molecular
mechanics and quantum chemistry calculations.  In one approach, a direct diagonalization of the
molecular mechanics Hessian
yields phonon frequencies $\omega_k$ and displacement vectors, and quantum chemistry calculations along
these displacements produce the coupling constants $c_k$.  Such an approach has been adopted recently
by Girlando et al.~in studies of electron and hole transport in rubrene\cite{gir10} and pentacene\cite{gir11}
crystals.

Alternatively, by appealing to the quantum-classical correspondence of
harmonic oscillators, which are presumed to compose the nuclear bath, one may show that the spectral
density can be obtained from the Fourier cosine transform of a classical 
correlation function\cite{cha91,ego99},
\begin{equation}
J_{ij}(\omega) = \frac{\omega}{k_B T} \int_0^\infty C^{cl}_{ij}(t) \cos(\omega t)
\end{equation}
where
$C^{cl}_{ii}(t) = \langle \delta E_i(t) \delta E_i(0) \rangle^{cl}_T$
is the energy gap fluctuation correlation function and
$C^{cl}_{i\neq j}(t) = \langle \delta V_{ij}(t) \delta V_{ij}(0) \rangle^{cl}_T$
is the electronic coupling fluctuation correlation function ($\delta X = X - \langle X \rangle_T^{cl}$).
This latter approach has been extensively pursued in the present context of organic materials by Troisi
and co-workers, who have focused on the fluctuations and spectral properties of the electronic coupling
in DNA\cite{tro02}, pentacene crystals\cite{tro06_jpca}, and the discotic phase of
hexabenzocoronene derivatives\cite{tro09}.

Given the expense of accurate \textit{ab initio} quantum chemistry methods, frequent calculations along
the course of a molecular dynamics trajectory are clearly prohibitive.  As such, it is common to adopt
a semi-empirical quantum chemical method, such as the spectroscopic parametrization of INDO (intermediate
neglect of differential overlap), which has an impressive accuracy to cost ratio allowing for
the rapid collection of sufficient statistics.  While one could in principle calculate all the diabatic
matrix elements defined in App.~\ref{app:basis}, we note that the diagonal elements are dominated by the bare orbital energies
and the off-diagonal coupling matrix elements are dominated by the one-electron coupling.  Thus it is
reasonable to assume that the stochastic properties (fluctuation magnitude and timescale) of
the full matrix element are equivalent to those
of its one-electron terms.  These latter properties are more commonly evaluated in the literature, due to
their role in electron and hole transport of organic materials.

The last topic of discussion concerns the correlation of different bath modes, for example the extent to
which the fluctuations of the diabatic energy of state $i$ are correlated with those of state $j$.
Although there is surely some degree of correlation, positive or negative, 
the effect of its inclusion on the subsequent dynamics is debatable.  In particular,
while some studies have attempted to implicate correlated bath modes in efficient biological energy transport\cite{str11,huo12},
molecular dynamics simulations of photosynthetic complexes show no significant correlations\cite{olb11,shi12}.
Lacking any firm evidence either way for the problem of singlet fission, 
we will let the correlation of different bath modes be dictated by numerical convenience (usually
preferring the completely uncorrelated scenario), though it is a topic worthy of further investigation.

\subsection{Reduced density matrix dynamics}

The dynamics of the coupled electron system and phonon bath is given by the Liouville-von Neumann equation
for the total density matrix, $W(t)$,
\begin{equation}
\frac{d W(t)}{d t} = -i \left[ \hat{H}_{tot}, W(t) \right],
\end{equation}
the exact solution of which is prohibitively difficult due to the large
Hilbert space associated with the phonon degrees of freedom.  However, as long as one is only interested
in electronic observables, great simplification occurs when considering the
\textit{reduced} density matrix (RDM) of the system, $\rho(t)$, obtained by averaging the total density matrix
over the phonon degrees of freedom, i.e. $\rho(t) = {\rm Tr}_{ph}\{W(t)\}$.
The diagonal elements of this matrix, $\rho_{ii}(t) = \langle i | \rho(t) | i \rangle$, are the populations
of state $i$ 
and the off-diagonal elements, $\rho_{ij}(t) = \langle i | \rho(t) | j \rangle$,
are the coherences between states $i$ and $j$.

A variety of methods exist for the determination of the RDM, each with its own caveats.
Although impressive progress has been made in the development of numerically exact methods -- including
path-integral techniques\cite{mak89,egg94,mak92,mak94}, the multi-configurational time-dependent Hartree ansatz\cite{mey90,man92,bec00},
and hierarchical equations of motion\cite{tan89,ish09_jcp2} -- we will limit ourselves here to approximate
methods which are more physically transparent and more readily applied to very large systems, as will be demonstrated
in our future work on clusters and crystals. 

Approximate methods are generally perturbative in nature, and differ in their choice of perturbative parameter.  Clearly,
the physical problem at hand should dictate the appropriate small parameter, thus controlling the accuracy of the 
perturbative approximation.  The first common approach is to treat the electronic couplings in the diabatic basis, $V_{ij}$,
to second order in perturbation theory, while treating the system-bath interaction exactly; this philosophy comprises
Marcus\cite{mar64} and F{\" o}rster-Dexter\cite{for53,dex53} theories, as well as the more sophisticated noninteracting blip approximation\cite{leg87,wei08}.  Although
this methodology has been previously employed in a study of CT-mediated singlet fission\cite{tei12}, to be discussed
later in this paper, we will advocate for an alternative
approach which treats the electronic couplings exactly in exchange for a perturbative treatment of the system-bath interaction.
The relative merits of the two approaches will be contrasted in Sec.~\ref{sec:redfield}.

Specifically, we shall pursue the use of a Redfield-like equation\cite{red65,blu81,pol96,bre02,ish09_jcp1}, in either its non-Markovian or
Markovian form.  Non-Markovian prescriptions can either take a time-local or time-nonlocal form, which corresponds to a
series resummation in terms of different time-ordered cumulants\cite{yoo75,muk78,muk78_cp}.  We will present equations for the time-local form (or partial ordering
prescription), though treatment of singlet fission dynamics in terms of the alternative time-nonlocal form (or complete
ordering prescription) would be straightforward.\footnote{It should be noted that the time-local approach should be favored in cases
where the system bath coupling is treated perturbatively, as in Sec.~\ref{sec:redfield}. This is because in the stochastic
limit, such a perturbation is Gaussian in nature, and thus partially ordered cumulants embody the correct statistics. An example
of this is seen in the discussion of pure dephasing in App.~\ref{app:spec}}

In the basis of electronic eigenstates, 
$\hat{H}_{el} |\alpha\rangle = \hbar \omega_\alpha |\alpha\rangle$,
and adopting the notation $\omega_{\alpha\beta} = \omega_\alpha-\omega_\beta$,
the time-local Redfield equation is given by
\begin{equation}\label{eq:redfield}
\frac{d\rho_{\alpha\beta}(t)}{dt} = -i\omega_{\alpha\beta}\rho_{\alpha\beta}(t)
	+ \sum_{\gamma,\delta} R_{\alpha\beta\gamma\delta}(t) \rho_{\gamma\delta}(t),
\end{equation}
where the initial condition of the total density matrix implicitly takes the factorized form
$W(0) = \rho(0) e^{-\hat{H}_{ph}/k_BT} / Z_{ph}$,
with the phonon partition function $Z_{ph} = {\rm Tr}_{ph} \{ e^{-\hat{H}_{ph}/k_BT} \}$.
This initial condition is consistent with an impulsive Franck-Condon excitation at time $t=0$.
In Eq.~(\ref{eq:redfield}), the first term
on the right hand side is responsible for coherent energy transfer whereas the
second term is responsible for population relaxation,
coherence transfer and dephasing, and more complicated population-to-coherence transfer processes.
Explicit expressions for the Redfield tensor elements, which include integrals over thermal bath correlation
functions, can be found in App.~\ref{app:redfield}.

In the limit where the bath relaxation takes place significantly faster than that of the electronic system,
the time-dependent Redfield equation is well approximated by its Markovian form, obtained from
Eq.~(\ref{eq:redfield}) by the replacement
$R_{\alpha\beta\gamma\delta}(t) \rightarrow R_{\alpha\beta\gamma\delta}(\infty)$.
This approximation clearly simplifies the form of the density matrix equation and provides a direct microscopic route
to dephasing and relaxation \textit{rates} which are often employed in other contexts as phenomenological parameters. 
This Markovian approximation should be carefully checked for its accuracy in each situation of interest.

As is commonly done in theories of exciton transport, one may furthermore
employ the secular approximation to the Markovian Redfield equation,
which preserves the positivity of the RDM, i.e. $\rho_{ii} > 0$\cite{blu81,bre02,may11,ish09_jcp1}.
The secular approximation amounts to neglecting those elements of the Redfield tensor, 
$R_{\alpha\beta\gamma\delta}$, for which $|\omega_{\alpha\beta}-\omega_{\gamma\delta}|\neq 0$.
In doing so, one decouples the dynamical evolution of populations
and coherences in the eigenstate basis.  In
addition to preserving positivity, the secular approximation furthermore guarantees that the system RDM approaches
thermal equilibrium at long times, i.e. 
\begin{equation}
\rho(t\rightarrow \infty) = e^{-\hat{H}_{el}/k_B T} / Z_{el},
\end{equation}
which is the correct physical result outside regimes of strong system-bath coupling.

\section{Applicability and accuracy of Redfield theory for singlet fission dynamics}\label{sec:redfield}

Although the presentation of system-bath dynamics up to this point has been largely generic,
we now thoroughly discuss the applicability of
the Redfield equation to the specific problem of singlet fission in organic systems.  We must first acknowledge the potential
disadvantages of the Redfield treatment and the extent to which they affect the reliability of such
calculations.  The main approximation inherent in this approach is the assumption of weak coupling
between the electronic and vibrational degrees of freedom.  This coupling can be quantified
approximately by the ratio of the \textit{magnitude} of fluctuations in the nuclei, $\lambda_{ij}$, to the 
\textit{frequency} of these fluctuations, $\Omega_{ij}$.  If the dimensionless ratio $\lambda_{ij} / \hbar \Omega_{ij}$
is small, then the Redfield approximation should be a good one.  Whether this inequality holds or not will depend
on the specific system under study.

As a prototypical singlet fission material, consider pentacene, which we will study in the follow papers.
The diagonal reorganization energy and
frequencies of the electron-phonon coupling have been calculated by quantum chemical and
molecular dynamics methods\cite{gir11} to be approximately 50 meV and 170 meV, respectively; note that this
latter value corresponds to the well-known $\approx$ 1400 cm$^{-1}$ aromatic stretching mode.  Thus we
see that the ratio of the two is indeed significantly smaller than one, and the Redfield equation should be
reasonably accurate.  As a general rule, smaller molecules will undergo larger geometry distortions in excited states,
i.e. larger $\lambda_{ii}$, and therefore the Redfield approach may break down.

The Markov approximation to the Redfield equation, as discussed above, relies on timescale
separation between electronic and nuclear relaxation, and thus one must compare the electronic frequencies $\omega_{ij}$ to those of the vibrations
$\Omega_{ij}$.  With electronic frequencies on the order of 50 meV, and again vibrational frequencies of 170 meV, even the Markov approximation
should be reasonably reliable.  In addition to this mathematical argument, there is also a more physical implication
of the Markov approximation: although the time-dependent variants of the Redfield equation can describe
multi-phonon effects to varying degrees of accuracy, the Markov approximation inherently describes only single-phonon
relaxation mechanisms.  This deficiency can be readily seen in the adiabatic population
relaxation rate, $R_{\alpha\alpha\beta\beta}$,
which is proportional to $J_{ij}(\omega_{\alpha\beta})$, so that all transition frequencies must be matched by a single phonon
frequency in the spectral density.

Potential pitfalls behind us, we now enumerate the many advantages of the Redfield formalism.  The first obvious advantage,
which is shared by a variety of other perturbative methods, is the clear microscopic formalism.  While density matrix
calculations have been employed for theoretical studies of MEG\cite{sha06} and singlet fission\cite{gre10_jpcb1}
as well as for fitting experimental singlet fission
data\cite{cha11}, such dynamical investigations have been essentially phenomenological to date. In the approach advocated here, 
the electronic structure methodology is directly connected to the molecular structure and microscopic relaxation mechanisms.  The Redfield tensor prescribes
temperature-dependent population relaxation and coherence dephasing rates which can be traced back to the physical
vibrations of the system under study.  When necessary, the time-dependent Redfield variants even yield non-Markovian behavior
which of course cannot be captured with phenomenological time-independent rates.  

In addition, like all master equation methods, the Redfield approach scales very favorably in a computational sense.  The additional adoption
of the Markov and secular approximations further reduces the computational cost.  Needless to say, none of the numerically exact methods
alluded to above takes on a simple master equation form and thus each has a significant computational overhead with a scaling that depends
on the details of the method.

The theoretical study most similar in
spirit to our own is that of Teichen and Eaves\cite{tei12} who sought to quantify the effects of a generally non-Markovian
bath of low-frequency solvent degrees of freedom and its implications for $CT$-mediated singlet fission.  These authors
employed methodology similar to the noninteracting blip approximation (NIBA) known from spin-boson theory\cite{leg87,wei08}, previously
generalized to the case of multilevel systems\cite{egg94,gol01} and recently extended to situations of slow, near-classical bath
modes\cite{ber12,ber12_eet} in a time-nonlocal formalism.  The time-nonlocal methodology, henceforth referred to as NIBA even for multilevel systems,
yields a non-Markovian master equation for the \textit{populations} of the RDM \textit{in the diabatic basis},
$P_i(t) = \rho_{ii}(t)$,
\begin{equation}
\frac{dP_i(t)}{dt} = \sum_j \int_0^t ds K_{ij}(t,s) P_j(s),
\end{equation}
where
\begin{equation}
K_{ij}(t,s) = 2 |V_{ij}|^2 \textrm{Re} \left\langle \exp\left({-i\hat{H}^{tot}_{ii}t}\right) \exp\left({i\hat{H}^{tot}_{jj}s}\right) \right\rangle_{ph},
\end{equation}
and $\hat{H}^{tot}_{ii} = \langle i | \hat{H}^{tot} | i\rangle$.
Teichen and Eaves instead considered the time-local version of this theory,
$\dot{P}_{i}(t) = I_i(t) + \sum_j R_{ij}(t) P_j(t)$, but the two methods should give similar results, and are identical
in the Markovian limit.

As alluded to previously, the NIBA-type master equations are perturbative in the electronic
couplings, $V_{ij}$, and thus the diabatic basis is in some sense a preferred basis.  The nonperturbative effects
of strong electronic coupling, yielding significant mixing in the adiabatic basis, cannot be described by the NIBA theory.
Accordingly, as a theory for populations only, NIBA makes no prediction about coherence variables, $\rho_{ij}(t)$,
preventing the transformation to any other electronic basis.  As described in more detail in App.~\ref{app:spec}, spectroscopy probes
the dynamics of coherences in the adiabatic basis, and as such is completely beyond reach of NIBA-based theories.

On the contrary, the nonperturbative nature of Redfield theory with respect to the electronic Hamiltonian allows for an
exact solution of the electronic structure problem in exchange for an approximate treatment of the system-bath interaction.
Thus all questions concerning delocalization, quantum coherence, and spectroscopy are readily addressed with the
Redfield framework, as long as the system-bath coupling is not too large.  Even in regimes where the time-dependent Redfield
theory is pushed past its limits of validity, the secular and Markovian approximations yield a numerically stable
Lindblad-type master equation, with microscopically-derived relaxation and coherence dephasing rates.  Interesting recent
work has formulated a stable theory which reinserts microscopic expressions for the population and coherence coupling
within the Lindblad formalism\cite{pal09}.

\subsection{Results for population dynamics}

Given the advantages of a Redfield-type approach with respect to the flexibility of treating populations and coherences on equal
footing in either the diabatic or adiabatic bases, as well as the ability to treat extremely large systems, it is natural
to ask if such an approach is accurate for typical singlet fission systems of current interest. Here we show with small
model systems that indeed treating the system-bath coupling as a perturbative parameter should yield semi-quantitative accuracy over
a wide range of scenarios rooted physically in the expected parameter space of acene systems.
In all of the following results on diabatic population dynamics, we make comparison with the numerically exact but computationally expensive hierarchical
equations of motion (HEOM) methodology\cite{tan89,ish09_jcp2,str12}, as implemented in the Parallel Hierarchy
Integrator ({\tt PHI}).\footnote{{\tt PHI}\cite{str12} was developed by the Theoretical and Computational Biophysics Group in
the Beckman Institute for Advanced Science and Technology at the University
 of Illinois at Urbana-Champaign. {\tt http://www.ks.uiuc.edu/Research/phi/}} To achieve convergence, we truncated the hierarchy
 at $L = 5$ and required $K = 3$ terms in the Matsubara expansion.

We begin with a two-state system, which in the context of singlet fission may be taken as a model for the direct, Coulomb-mediated
fission mechanism.  The first state is the photoexcited initial singlet, $S_1$, and the second state is the
multi-exciton configuration, $TT$.  The initial condition is $\rho(0) = |S_1\rangle\langle S_1|$ and the dynamics proceeds
based on the parameters of the system-bath Hamiltonian defined above.  The system-bath coupling will be chosen to take the
simple form $H_{el-ph} = \sum_{i=S_1,TT }\sum_{k,i} |i\rangle c_{k,i} q_{k,i} \langle i|$, i.e. linear, diagonal coupling to
uncorrelated bath degrees of freedom.  The baths will be characterized by identical Ohmic spectral densities with a Lorentzian cutoff
(sometimes referred to as the overdamped Brownian oscillator model), $J_{ii}(\omega) = 2\lambda\Omega \omega / (\omega^2 + \Omega^2)$.

\begin{figure}[t]
\centering
\includegraphics[scale=1.0]{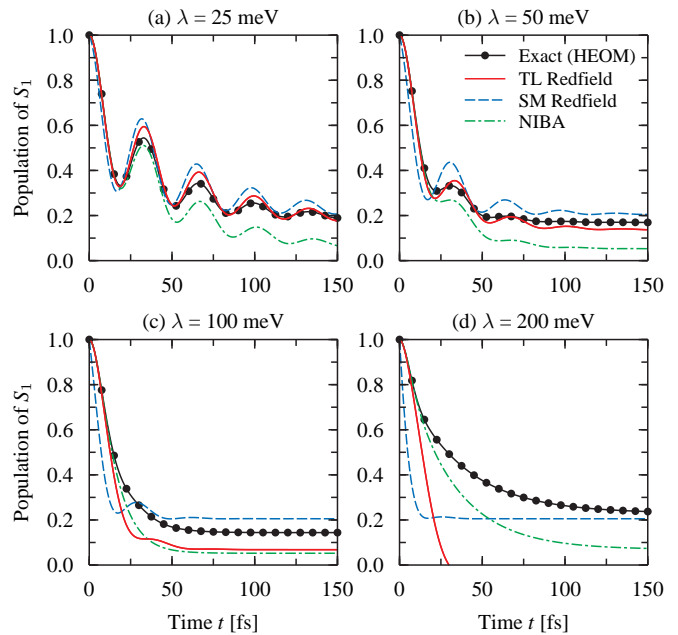}
\caption{
Population dynamics of the two-state singlet fission model described in the text with $E_{S_1}-E_{TT} = 75$ meV, $V = 50$ meV,
$\hbar \Omega = 150$ meV ($\Omega^{-1} \approx$ 4 fs), and $T$ = 300 K ($k_BT \approx 26$ meV), for increasing system-bath coupling strength.
}
\label{fig:2a}
\end{figure}

For concreteness, we will fix the fission to be
mildly exothermic, $E_{S_1} - E_{TT} = 75$ meV, with a bath cutoff frequency $\hbar \Omega = 150$ meV (characteristic of aromatic molecules)
and temperature $T = 300$ K ($k_BT \approx 26$ meV).
However, to investigate the perturbative accuracy of the Redfield and NIBA equations, we will scan the reorganization
energy, $\lambda$, and electronic coupling, $V$.

In Fig.~\ref{fig:2a}, we plot the singlet population dynamics for a variety of reorganization energies, with the electronic
coupling fixed at $V = 50$ meV.  When the system-bath coupling is weak, the quantum beating is dominant and overall relaxation
is slow. It is clear that the timescale of beating should not be confused with the relaxation timescale; only the former
is accessible within static electronic structure calculations, whereas the latter requires explicit treatment of the vibrational
degrees of freedom.  As the coupling is increased, all theories correctly predict that the oscillations become damped
and the relaxation to $TT$ proceeds more quickly.  We see that as the system-bath interaction becomes large, the time-local
Redfield result becomes inaccurate at long times, even leading to unphysical negative populations.  However, to some extent,
the secular and Markov approximations to the Redfield equation prevent such a catastrophe, leading to much more reasonable
equilibrium populations.  The non-Markovian behavior, on the other hand, can be observed in the short-time dynamics, which
are always correctly described by the time-local Redfield equation, but not by its secular, Markovian counterpart.  In contrast
to the breakdown behavior of the Redfield equation, the NIBA dynamics retain their relative accuracy at all values of the
system-bath coupling.  This result is to be expected in as much as the NIBA theory treats the system-bath interaction exactly.
Rather, the NIBA theory is perturbative in the electronic coupling, which is unchanged in all panels of Fig.~\ref{fig:2a}.
However, NIBA can be seen to consistently underestimate the equilibrium population of $S_1$.  This tendency towards
extreme localization in biased systems is a known deficiency of methods that are perturbative in the electronic coupling\cite{mak95,leg87,wei08}.

\begin{figure}[t]
\centering
\includegraphics[scale=1.0]{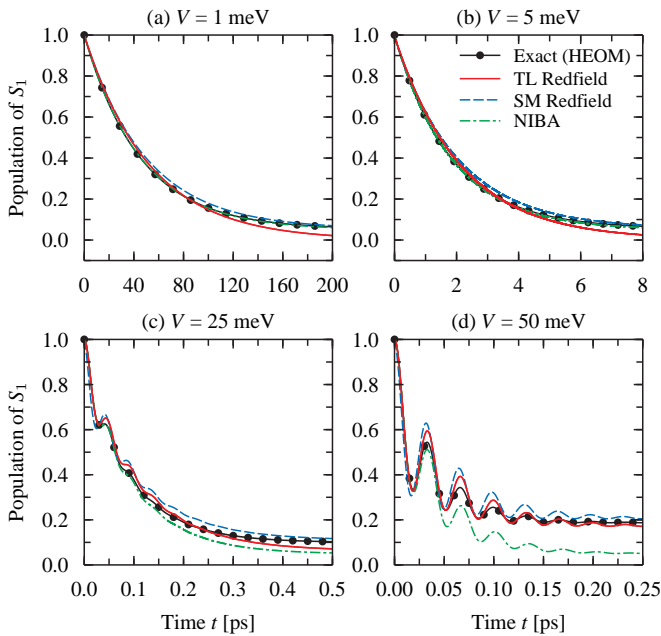}
\caption{
The same as in Fig.~\ref{fig:2a}, but with $\lambda = 25$ meV and scanning the electronic coupling, $V$.  Note the changing
scale of the time axis.
}
\label{fig:scanj}
\end{figure}

The rapid singlet fission observed in Fig.~\ref{fig:2a}, with 100 fs timescales, is due to the rather large value of the
direct coupling matrix element, $V$.  As alluded to previously, this number is likely significantly smaller than 50 meV and so we show,
in Fig.~\ref{fig:scanj}, the effect of reducing the electronic coupling.  As $V$ gets progressively smaller, the timescale
of relaxation grows significantly, reaching approximately 100 ps for $V$ = 1 meV.  At the smallest values of $V$, panels (a)
and (b), the relaxation rate can be seen to follow the expected $k \propto V^2$ golden rule.  Consistent with their perturbative
origins, the NIBA dynamics become quantitatively exact for vanishing $V$, whereas Redfield theory's qualitative accuracy is
maintained throughout all panels.  Interestingly, for this value of the reorganization energy, the secular and Markov approximations
to the time-local Redfield equation yield impressive quantitative accuracy for all values of $V$.

As another important numerical test, we now consider the effect of a third state on the dynamics of singlet fission,
where an initial state couples to a second state which is in turn coupled to a third.  This configuration
is clearly akin to the mediated mechanism, with the three states $S_1$, $CT$, and $TT$.  Interestingly, the quantum
dynamics of such mediating systems has precedent in the donor-bridge-acceptor complexes of photosynthetic charge transfer.
Over 15 years ago, Makri and coworkers performed numerically exact path integral 
simulations of a three-state model very similar
to the one considered here and detailed the dynamical features of
two previously proposed, but physically distinct transport mechanisms\cite{mak96,sim97}.  The first mechanism is a \textit{sequential}
one, whereby the intermediate state becomes transiently populated in a scheme well-described by two-step kinetics. This
mechanism dominates in energetic regimes satisfying $E_1 > E_2 > E_3$.  A second mechanism, evincing the \textit{superexchange}
phenomenon, employs virtual states of the intermediate which is never directly populated, yielding overall single-step
kinetics for the $1 \rightarrow 3$ transfer.  Superexchange applies when the intermediate state energy is much higher
than the other two, $E_2 \gg E_1 > E_3$.

In light of the similarity with mediated singlet fission (both qualitatively and quantitatively, see below), we consider it an essential
criterion that any adopted quantum
dynamics methodology be able to capture this effect. 
To make connection with the singlet fission problem, we will henceforth consider the definite state labeling referred to
above, namely $S_1$, $CT$, and $TT$.
For comparison, we adopt the same electronic parameters used in
Ref.~\onlinecite{mak96} and the same system-bath coupling used above,
$\hat{H}_{el-ph} = \sum_{i=S_1,CT,TT} |i\rangle \sum_{k,i} c_{k,i} q_{k,i} \langle i |$.
Again the baths have an Ohmic spectral density with
Lorentzian cutoff parametrized by $\lambda = 25$ meV, $\hbar \Omega = 150$ meV, and $T = 300$ K.
We wish to strongly emphasize that although this Hamiltonian (electronic parameters to follow)
was originally parametrized based on photosynthetic protein data, the magnitude
of the parameters is almost identical with those expected of singlet fission.

\begin{figure}[t]
\centering
\includegraphics[scale=1.0]{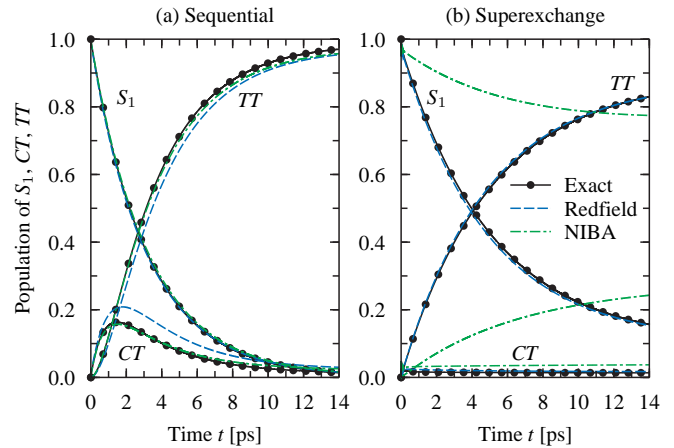}
\caption{
Population dynamics of the three-state model described in the text. While both Redfield theory and NIBA provide a reliable description
of dynamics in the two-step sequential regime, NIBA is qualitatively
unable to describe the superexchange regime. Redfield equation dynamics employ the secular and Markov approximations and exact results
are calculated with the HEOM approach.
}
\label{fig:super}
\end{figure}

First, we consider the sequential regime, for which the electronic Hamiltonian has
$E_{S_1} = 0$, $E_{CT} = -50$ meV, and $E_{TT} = -250$ meV, with $V_{S_1,CT}=3$ meV and
$V_{CT,TT} = 17$ meV. In Fig.~\ref{fig:super}(a), we plot the population dynamics of the three states and find
that both Redfield theory and NIBA agree quantitatively with each other and with numerically exact HEOM dynamics.
Clearly, in this sequential mechanism,
$S_1$ populates $CT$, which in turn populates $TT$.

We now turn to the superexchange regime, for which $E_{S_1} = 0$, $E_{CT} = 250$ meV, and $E_{TT} = -80$ meV, with $V_{S_1,CT} = V_{CT,TT} = 30$ meV;
note that the immense barrier, $E_{CT} - E_{S_1} = 250$ meV $ \approx 10 k_B T$ completely prohibits thermal activation.
Turning to the results in Fig.~\ref{fig:super}, the Redfield and NIBA dynamics strongly differ, and only Redfield theory gives results in good quantitative agreement
with the exact dynamics. In the superexchange limit, the intermediate $CT$ state is never directly populated, and the kinetics follows
a simple $S_1 \rightarrow TT$ dynamical scheme.  Thus, in spite of the relatively small electronic coupling values, $V_{S_1,CT} = V_{CT,TT} = 30$ meV, superexchange
must be understood as a higher-order effect. The effective electronic coupling from $S_1$ to $TT$
due to $CT$-mixing is $V_{\rm eff} \approx V_{S_1,CT} V_{CT,TT} / (E_{S_1} - E_{TT})$. Performing second-order perturbation theory with
this effective electronic coupling thus gives a rate which is overall fourth-order, explaining why superexchange eludes the usual second-order treatment,
such as that employed in NIBA.  Redfield theory, on the other hand, is completely nonperturbative in the electronic couplings, and is thus entirely
capable of capturing this highly relevant phenomenon.
Although we will have more to say about it in our future work, this simple model clearly refutes arguments that high-lying $CT$ intermediates preclude
efficient mediated singlet fission.

\subsection{Results for spectroscopy}

\begin{figure}[t]
\centering
\includegraphics[scale=1.0]{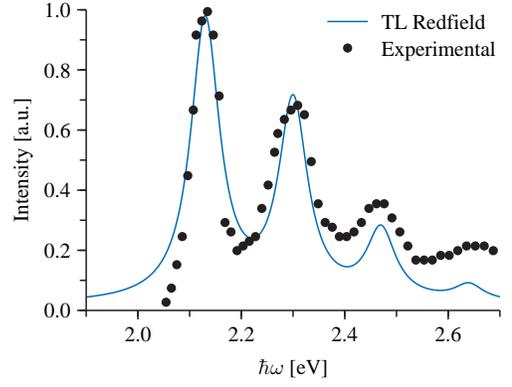}
\caption{
Calculated and experimental absorption spectrum of a single pentacene molecule at $T=100$ K in solution.  The
calculation parameters are $E(S_1) = 2.3$ eV, $\Omega = 170$ meV, and $\lambda = 120$ meV ($S = 0.7$),
and a homogeneous broadening of 30 meV has been applied.  Experimental spectrum is from Ref.~\onlinecite{tao07}.
}
\label{fig:abs_pc}
\end{figure}

Lastly, we apply the Redfield formalism to the calculation of linear absorption spectroscopy, the formalism of which is described in App.~\ref{app:spec}.
In this situation, the non-Markovian time-local
variant is to be preferred, as it exactly solves the so-called pure-dephasing problem appropriate for the single-molecule absorption spectrum.
Employing the methodology described there, we have calculated the absorption of a pentacene
molecule in solution, which compares very favorably with the experimental spectrum, see Fig.~\ref{fig:abs_pc}.  In particular, the phonon sidebands are accurately
reproduced, even though this is a purely non-Markovian, multi-phonon signature.  As such, this feature cannot be described by the Markovian
Redfield theory.  Comparisons such as this one provide essential benchmarks for the accurate parametrization of both electron and phonon degrees
of freedom, as well as the interaction between them.

In the presence of intermolecular interactions, the absorption lineshape will be changed from that of Fig.~\ref{fig:abs_pc} due to the
electronic coupling to other excited states.
To discuss these effects, let us introduce the dipole operator which, to a reasonable approximation, is given by
\begin{equation}
\hat{\mu} = \sum_m |(S_1)_m\rangle \bm{\mu}_{S_1} \langle S_0 | + {\rm H.c.},
\end{equation}
where H.c.~denotes the Hermitian conjugate.  This approximation follows from the observation that the transition
dipole matrix element for a single excitation from molecule $m$ to molecule $i$,
\begin{equation}
\bm{\mu}_m^i \equiv -e \sum_n \langle 0 | \bm{r}_n | \Psi_m^i \rangle = -e \left( H_m | \vr | L_i \right),
\end{equation}
is significantly larger for $m=i$ (Frenkel excitations) than for $m\neq i$ (CT excitations), due to
spatial locality.
Thus
only the intramolecular Frenkel excitation has a non-negligible transition dipole moment, and charge-transfer
and multiple-exciton states do not absorb light.  However, the expression for the absorption signal (see Eq.~(\ref{eq:abs}))
requires the \textit{adiabatic} transition dipole
moments, which follow from the above equation and the diagonalizing transformation, $|\alpha\rangle = \sum_i C_i^\alpha |i\rangle$, as
\begin{equation}
\bm{\mu}_{\alpha} = \langle 0 | \hat{\mu} | \alpha \rangle \approx \sum_{(S_1)_n} C^\alpha_{(S_1)_n} \bm{\mu}_{S_1}.
\end{equation}
It is crucially important to recognize that the extent to which the adiabatic state $\alpha$ is composed of
diabatic intramolecular excitations ($C_{(S_1)_n}^\alpha$) determines the strength with which it absorbs; this is the phenomenon of
\textit{intensity borrowing}.  Therefore, signatures of ``dark'' diabatic states, such as charge-transfer or multi-exciton states,
will appear in absorption spectra if these states are strongly coupled to the ``bright'' intramolecular
Frenkel excitations.

This phenomenon has been previously addressed in the MEG literature within the context of the coherent superposition
approximation, wherein the authors concluded that coupling to multi-exciton states does not affect the total
absorption coefficient, $\alpha$\cite{sha06}.  While we agree that the \textit{integrated} absorption coefficient is indeed only determined
by the bright singlet excitons,
\begin{equation}
\begin{split}
\alpha &= \int_0^\infty d\omega I(\omega)
    \approx \sum_\alpha |\mu_{\alpha,\epsilon}|^2\ \int_0^\infty d\omega \delta(\omega - \omega_{\alpha0}) \\
    &= \sum_\alpha |\mu_{\alpha}|^2 = \sum_\alpha \sum_{(S_1)_m} \sum_{(S_1)_n}
        \left[C_{(S_1)_m}^\alpha\right]^* C_{(S_1)_n}^\alpha |\mu_{S_1}|^2 \\
    &= \sum_{(S_1)_m} \sum_{(S_1)_n} \langle (S_1)_m | (S_1)_n \rangle |\mu_{S_1}|^2 = N |\mu_{S_1}|^2,
\end{split}
\end{equation}
we argue that the spectral \textit{structure} (peak positions and intensities) is surely affected by the coupling to multi-exciton states.
For example, consider an artificial system composed of the ground state, $|0\rangle$,
a single bright state, $|S_1\rangle$, and a dark (multi-exciton) state, $|D\rangle$,
with equal energies $E(S_1) = E(D) \equiv E$ and mutual coupling $\langle S_1 | \hat{H}_{el} | D \rangle \equiv V$.  The dipole operator
is simply $\hat{\mu} = \mu |S_1\rangle\langle 0| + {\rm H.c.}$  In the uncoupled limit $V=0$, the electronic Hamiltonian
is already diagonal, and Eq.~(\ref{eq:abs_el}) gives the absorption lineshape as 
$I(\omega) = |\mu|^2 \delta(\omega-E)$, and the absorption coefficient $\alpha = |\mu|^2$.  However, for
$V \neq 0$, the adiabatic eigenstates are the symmetric and antisymmetric linear combinations,
$|\alpha\rangle = (2)^{-1/2} (|S_1\rangle + |D\rangle)$, $|\beta\rangle = (2)^{-1/2} (|S_1\rangle - |D\rangle)$,
with energies $E_\alpha = E-V$, $E_\beta = E+V$.  In this case, the absorption lineshape shows two weaker peaks,
\begin{equation}
I(\omega) = \frac{1}{2} |\mu|^2 \left[ \delta(\omega - E + V) + \delta(\omega - E - V) \right]
\end{equation}
but the same total absorption coefficient, $\alpha = |\mu|^2$.
In light of the preceding analysis, we propose that evidence for coupling to dark, perhaps multi-exciton, states should
be observable in the linear absorption spectrum, perhaps contrary to standard intuition.

As a related point, we recall that there have been proposals 
to experimentally seek out real-time quantum beating as evidence of coupling to multi-exciton states\cite{sha06},
even if, in the limit of strong coupling,
the frequency of beating becomes too high for experimental resolution in the time domain.  
However, the origin of spectral peaks discussed above is identical to that of quantum beating, namely the oscillation
of quantum coherences at frequencies given by energy differences.  More importantly, in the frequency domain, the
peak splitting is proportional to the strength of the coupling and thus more easily observed for strong coupling.
In this sense, linear absorption and real-time quantum beating should be seen as complementary tools for the investigation
of coupling to multi-exciton states: weak coupling yields negligibly split peaks that may be difficult to resolve, but produces 
slow oscillations in real time that should be easy to observe.  The situation is reversed for strong
coupling, where spectral measurements should be preferred.  As a proviso, the real-time observation of quantum beating
may be an artifact of the diabatic basis.  Specifically, the adiabatic populations may show pure exponential relaxation,
but because the transformation back to the diabatic basis mixes populations with oscillatory coherences, the diabatic
populations appear to exhibit quantum beating.  Thus the real-time detection of such beating in part depends on the
basis which is probed experimentally.

The examples discussed in this section on spectroscopy and the preceding one on population dynamics
illustrate the utility of a Redfield approach to the description of organic
singlet fission systems. In particular, most materials of current interest for singlet fission lie in a regime where the ratio
$\lambda_{ii}/\hbar \Omega_{ii} < 1$, largely due to the dominant coupling to high-frequency aromatic carbon bond stretching.
For the same reason, these systems generically have bath relaxation times that make the simplifying Markov approximation a sensible one.
Lastly, Redfield theory and its variants are non-perturbative in the electronic states.  Thus, they do not alter the underlying
description of the frozen electronic structure theory and are capable of treating higher-order effects such as superexchange.
This last point will be significant in our discussion of singlet fission in pentacene dimers.

\section{Conclusions}\label{sec:conc}

To summarize, we have presented a self-contained microscopic theory of multi-exciton formation in the
context of singlet fission.  Our formalism emphasizes the difference in electronic bases, diabatic
and adiabatic, as applied to both theoretical development and experimental interpretation. Building
on this electronic foundation, we have applied techniques from the theory of open quantum systems
to describe the finite-temperature quantum dynamics of charge and energy transfer processes taking
place in singlet fission materials.  Specifically, such processes are facilitated by phonon absorption
and emission, which can be given a microscopic origin in terms of certain vibrational motions of molecules.

We furthermore discussed various approximate quantum master equations for the reduced density matrix
and found that while NIBA-like theories which are perturbative in the electronic couplings yield
accurate dynamics for two-state systems in regimes expected to hold in organic systems undergoing singlet
fission, their perturbative nature is exposed by higher-order processes,
such as the superexchange mechanism.  On the contrary, Redfield-like theories,
perturbative in the system-bath coupling, yield reasonably accurate results in essentially all
regimes of relevance
for singlet fission.  We additionally elucidated the importance of a theory for both the populations
and coherences of the reduced density matrix, allowing for investigation of dynamics in different
electronic bases as well as prediction of linear spectroscopies such as absorption.  While more accurate
quantum dynamics scheme are certainly worthy of consideration, we emphasize the efficiency of
Redfield theory, which allows for a rapid, but thorough, investigation of parameter space in both
small and large systems, which is ideal for a computational screening of efficient singlet fission target materials.

In the small model systems considered here, we have found that both direct and mediated mechanisms are plausible
pathways to efficient singlet fission.  For reasonable electronic Hamiltonian parameters, the phonon
degrees of freedom facilitate fission on picosecond timescales.  In particular, we drew a potentially useful
comparison with charge transport in photosynthetic donor-bridge-acceptor systems in the context of
$CT$-mediated fission, wherein both sequential and superexchange mechanisms should be considered possible.
We emphasize that our proposal for a unified framework for the microscopic treatment of singlet fission in organic
systems is based on accuracy, efficiency, and physical transparency.  In particular,
we have generalized existing techniques and used physical arguments and numerical benchmarks
to marry them for the purpose of a microscopic and accurate treatment of fission in systems ranging
from dimers to crystals.  The companion paper to this one begins this program in pentacene dimers
while future work will consider large aggregates and bulk crystals.

\begin{acknowledgments}
This work was supported in part by the Center for Re-Defining Photovoltaic Efficiency through Molecule Scale
Control, an Energy Frontier Research Center funded by the US Department of Energy, Office of Science, Office of
Basic Energy Sciences under Award Number DE-SC0001085.
This work was carried out in part at the Center for Functional Nanomaterials, Brookhaven National Laboratory,
which is supported by the U.S. Department of Energy, Office of Basic Energy Sciences under Contract No. DE-AC02-98CH10886 (M.S.H).
T.C.B. was supported in part by the Department of Energy Office of Science Graduate Fellowship Program (DOE SCGF),
made possible in part by  the American Recovery and Reinvestment Act of 2009, administered by ORISE-ORAU
under contract no. DE-AC05-06OR23100.
\end{acknowledgments}

\appendix

\section{Energies and couplings in truncated CI basis}\label{app:basis}

Using the minimal basis presented in Sec.~\ref{ssec:minimal}, here we give formulas for the diagonal and off-diagonal matrix
elements of the exact electronic Hamiltonian operator,
\begin{equation}\label{eq:ham}
\hat{H}_{el} = \sum_{ij,\sigma} h_{ij} c^\dagger_{i,\sigma}c_{j,\sigma}
	+ \frac{1}{2}\sum_{ijkl,\sigma\sigma^\prime} V_{ijkl} c^\dagger_{i,\sigma}c^\dagger_{j,\sigma^\prime}
		c_{l,\sigma^\prime} c_{k,\sigma}
\end{equation}
where the sums are over \textit{all} molecular orbitals of the isolated molecules (the indices $i,j,k,l$ now contain
both the molecule and its orbital) and
\begin{equation}
h_{ij} = \int d^3\vr \phi^*_i(\vr) \left[ -\frac{1}{2}\nabla_{\vr}^2 + V_{el-nuc}(\vr) \right] \phi_j(\vr) \equiv( i | \hat{h} | j ),
\end{equation}
\begin{equation}
V_{ijkl} = \int d^3\vr_1 \int d^3\vr_2 \phi^*_i(\vr_1) \phi^*_j(\vr_2) r_{12}^{-1} \phi_k(\vr_1) \phi_l(\vr_2) \equiv ( ij | kl )
\end{equation}
are one- and two-electron integrals over \textit{spatial} orbitals.
Here and henceforth we employ atomic units.

The diagonal matrix elements of the Hamiltonian operator in the diabatic basis (``energies'') are given by
expressions available from known CI theory,
\begin{equation}
E[S_0] = \sum_{i\in S_0} 2(i | \hat{h} | i) + \sum_{i,j\in S_0} 2(ij | ij ) - (ij | ji)
\end{equation}
\begin{equation}
E\left[(S_1)_m\right] = E[S_0] + E_g + 2K_{H_m L_m} - J_{H_m L_m},
\end{equation}
\begin{equation}
E\left[C_mA_n\right] = E[S_0] + E_g + 2K_{H_m L_n} - J_{H_m L_n},
\end{equation}
and
\begin{equation}
\begin{split}
E\left[T_m T_n\right] &= E[S_0] + 2E_g + J_{H_m H_n} + J_{L_m L_n} \\
&\hspace{1em} - J_{L_m H_m} - J_{L_m H_n} - J_{L_n H_m} - J_{L_n H_n} \\
&\hspace{1em} - (1/2) \left( K_{H_m H_n} + K_{L_m L_n} \right) \\
&\hspace{1em} + \left[\left(6\sqrt{3} + 5\right)/4\right]\left( K_{L_m H_n} + K_{L_n H_m} \right) \\
&\hspace{1em} - \left[\left(6\sqrt{3} - 5\right)/4\right]\left( K_{L_m H_m} + K_{L_n H_n} \right).
\end{split}
\end{equation}
We have introduced the gap $E_g = \varepsilon_{L_m} - \varepsilon_{H_m}$ and notation for direct Coulomb integrals,
$J_{ij} = (ij | ij)$,
and exchange integrals,
$K_{ij} = (ij | ji)$.

While one could in principle calculate all possible couplings between the previously introduced 
diabatic states, those couplings involving three or four molecules will be negligibly small
due to the weak overlap of the localized molecular orbitals.
Specifically, we propose to neglect all three-center and higher two-electron integrals.
This semi-empirical approximation should not drastically affect the results.

Introducing the \textit{spatial} orbital matrix elements of the Fock operator, $\hat{F}$,
\begin{equation}
(i|\hat{F}|j) = (i | \hat{h} | j ) + \sum_{k\in S_0} 2(ik|jk) - (ij|kk),
\end{equation}
the remaining off-diagonal matrix elements (``couplings'') can be evaluated to give
\begin{align}
\begin{split}
\langle C_m A_n | \hat{H}_{el} | (S_1)_m \rangle &= ( L_m | \hat{F} | L_n ) \\
&\hspace{1em} + 2 (H_m L_m | L_n H_m ) - (H_m L_m | H_m L_n )
\end{split} \\
\begin{split}
\langle C_m A_n | \hat{H}_{el} | (S_1)_n \rangle &= -( H_m | \hat{F} | H_n ) \\
&\hspace{1em} + 2 (H_n L_n | L_n H_m ) - (H_n L_n | H_m L_n )
\end{split} \\
\langle C_m A_n | \hat{H}_{el} | C_n A_m \rangle &= 2 ( H_m L_m | L_n H_n ) - ( H_m L_m | H_n L_n ) \\
\begin{split}
\langle C_m A_n | \hat{H}_{el} | T_m T_n \rangle &= \sqrt{3/2} \Big\{ ( L_m | \hat{F} | H_n ) \\
&\hspace{1em} + (L_m L_n | H_n L_n ) - (L_m H_m | H_n H_m ) \Big\}
\end{split} \\
\langle (S_1)_m | \hat{H}_{el} | (S_1)_n \rangle &= 2 ( H_m L_n | L_m H_n ) - ( H_m L_n | H_n L_m ) \\
\langle (S_1)_m | \hat{H}_{el} | T_m T_n \rangle &= \sqrt{3/2} \Big\{ ( L_m L_n | H_n L_m ) \\
&\hspace{1em} - ( H_m H_n | L_n H_m ) \Big\}.
\end{align}

We have neglected the coupling to the ground state, e.g. $\langle C_m A_n | \hat{H}_{el} | S_0 \rangle$.
Note that such terms do not strictly vanish as they do in traditional CIS (Brillouin's theorem)
because the reference state is not the Hartee-Fock solution of the full molecular cluster.
Nonetheless, because the energy gap between ground and excited states is large, the renormalization
of excited states due to coupling to the ground state should be negligible.  Furthermore, although these
terms could in principle facilitate non-radiative decay to the ground state, we assume a bottleneck for phonon
emission prevents such events, justifying our neglect of such couplings.

\section{Derivation of the system-bath Hamiltonian}\label{app:sysbath}

We begin by considering the nuclear dependence of the diabatic electronic state energies
and couplings introduced above,
i.e. $U_i(\vQ) = \langle i | \hat{U}(\vQ) | i \rangle$ and $V_{ij}(\vQ) = \langle i | \hat{U}(\vQ) | j \rangle$.
To simplify notation, we will employ the Roman characters $i$ and $j$ to denote diabatic states and
Greek characters ($\alpha,\beta,\gamma,...$) to denote the adiabatic eigenstates.

In the limit where the ground diabatic state is a local minimum, we may perform a second-order Taylor expansion of the potential
in terms of the mass-weighted coordinates $\bar{\vQ}_n \equiv \sqrt{M_n}(\vQ_n - \vQ_n^{(0)})$ and
employ a transformation to the normal modes that diagonalize the
Hessian $q_k = \sum_{n=1}^N\sum_{x=1}^3 u_{n,x}^k \bar{Q}_{n,x}$, yielding
\begin{equation}
U_0(\{q_n\}) \approx E_0 + \frac{1}{2} \sum_{k} \omega_k^2 q_k^2,
\end{equation}
where $E_0 = U_0(\{0\})$ and $\{\omega_k^2\}$ are the eigenvalues of the Hessian\cite{may11}.
By promoting the nuclear coordinates to operators and re-inserting the kinetic energy operator,
we arrive at the ground state diagonal matrix element of an electron-phonon Hamiltonian in normal-mode coordinates,
\begin{equation}
\langle 0 | \hat{H}_{tot} | 0 \rangle = E_0 + \sum_{k} \left[ \frac{\hat{p}_k^2}{2} + \frac{1}{2} \omega_k^2 \hat{q}_k^2 \right].
\end{equation}
The matrix elements of the total Hamiltonian in the higher-lying diabatic states thus additionally acquire a linear term
describing the shift of the excited state potential energy surface minimum located at $\{q_k^{(i)}\}$,
\begin{equation}
\begin{split}
\langle i | \hat{H}_{tot} | i \rangle &= U_i(\{q_k^{(i)}\}) +  \sum_{k} \left[ \frac{\hat{p}_k^2}{2}
    + \frac{1}{2} \omega_k^2 \left(\hat{q}_k-q^{(i)}_k\right)^2 \right] \\
&\equiv E_i + \sum_{k} \left[ \frac{\hat{p}_k^2}{2} + \frac{1}{2} \omega_k^2 \hat{q}_k^2 + c_{k,i} \hat{q}_k\right],
\end{split}
\end{equation}
where the vertical energy is $E_i \equiv U_i(\{0\}) = U_i(\{q_k^{(i)}\}) + \lambda_{ii} $, the Holstein-like coupling constants
are given by\cite{hol59,gir11}
$c_{k,i} = - \omega_k^2 q_k^{(i)}$
and the reorganization energy\cite{bre02,may11} of state $i$ is defined as
\begin{equation}
\lambda_{ii} = \frac{1}{2} \sum_{k} \omega_k^2 \left[q_k^{(i)}\right]^2 = \frac{1}{2}\sum_{k} \frac{c_{k,i}^2}{\omega_k^2}.
\end{equation}
It will be convenient to now define the so-called \textit{spectral density} of the phonons, which completely characterizes
the harmonic environment intrinsic in the normal-mode decomposition.  The spectral density of state $i$ is defined by
\begin{equation}
J_{ii}(\omega) =  \frac{\pi}{2} \sum_{k} \frac{c_{k,i}^2}{\omega_k} \delta(\omega-\omega_k)
\end{equation}
from which the reorganization energy can be rewritten as
$\lambda_{ii} = \pi^{-1} \int_0^\infty d\omega J_{ii}(\omega)/\omega$.

The coordinate dependence of the off-diagonal matrix elements in these normal modes
can then also be evaluated to first order,
\begin{equation}
\langle i | \hat{H}_{tot} | j \rangle = V_{ij} + \sum_{k} c_{k,ij} q_k,
\end{equation}
with the Peierls-like coupling constants given by\cite{pei55,gir11}
\begin{equation}
c_{k,ij} = \frac{\partial V_{ij}(\{q_n\})}{\partial q_k} \Bigg|_{\{q_n\}=\{0\}}.
\end{equation}
These coupling constants can also be incorporated into an off-diagonal spectral density
\begin{equation}
J_{ij}(\omega) = \frac{\pi}{2} \sum_{k} \frac{c_{k,ij}^2}{\omega_k} \delta(\omega-\omega_k)
\end{equation}
with a corresponding ``reorganization'' energy
$\lambda_{ij} = \pi^{-1} \int_0^\infty d\omega J_{ij}(\omega)/\omega$.

Combining all of the above, we thus arrive at the desired system-bath-type Hamiltonian, Eqs.~\ref{eq:ham_sb}.

\section{Explicit expressions for the Redfield tensor elements}\label{app:redfield}

The Redfield tensor is given explicitly as\cite{bre02,ish09_jcp1}
\begin{equation}
\begin{split}
R_{\alpha\beta\gamma\delta}(t) &= \Gamma^+_{\delta\beta\alpha\gamma}(t) + \Gamma^-_{\delta\beta\alpha\gamma}(t) \\
	&\hspace{1em}-\delta_{\delta\beta}\sum_\kappa \Gamma^+_{\alpha\kappa\kappa\gamma}(t)
	-\delta_{\alpha\gamma}\sum_\kappa \Gamma^-_{\delta\kappa\kappa\beta}(t)
\end{split}
\end{equation}
where
\begin{align}
\Gamma^+_{\alpha\beta\gamma\delta}(t) &= 
\int_0^t ds 
e^{-i\omega_{\gamma\delta}s}
	\langle \hat{H}^{el-ph}_{\alpha\beta}(s)\hat{H}^{el-ph}_{\gamma\delta}(0) \rangle_{ph} \label{eq:gammaplus}\\
\Gamma^-_{\alpha\beta\gamma\delta}(t) &= 
\int_0^t ds
e^{-i\omega_{\alpha\beta}s}
	\langle \hat{H}^{el-ph}_{\alpha\beta}(0)\hat{H}^{el-ph}_{\gamma\delta}(s) \rangle_{ph} \label{eq:gammaminus}
\end{align}
are integrals of thermal bath correlation functions.  We have introduced the notation
\begin{equation}
\hat{H}^{el-ph}_{\alpha\beta}(t) \equiv \sum_{i,j} \langle \alpha | i\rangle
	e^{i\hat{H}_{ph} t/\hbar} \langle i | \hat{H}_{el-ph} | j\rangle e^{-i\hat{H}_{ph} t/\hbar} \langle j | \beta \rangle
\end{equation}
and $\langle \dots \rangle_{ph} \equiv {\rm Tr}_{ph} \{ \dots e^{-\hat{H}_{ph}/k_BT}/Z_{ph} \}$.
The calculation of the thermal bath correlation functions required for the Redfield relaxation tensor is
straightforward for the harmonic baths derived above.  Assuming uncorrelated bath fluctuations, one finds
\begin{equation}
\sum_{k,k^\prime} c_{k,i} c_{k^\prime,j}
    \langle \hat{q}_{k,i}(t) \hat{q}_{k^\prime,j}(0) \rangle_{ph} = \delta_{ij} C_{ii}(t),
\end{equation}
\begin{equation}
\sum_{k,k^\prime} c_{k,ij} c_{k^\prime,mn} \langle \hat{q}_{k,ij}(t) \hat{q}_{k^\prime,mn}(0) \rangle_{ph} 
	= (\delta_{im}\delta_{jn}+\delta_{in}\delta_{jn}) C_{ij}(t),
\end{equation}
and $\langle \hat{q}_{k,ij}(t) \hat{q}_{k^\prime,m}(0) \rangle_{ph} = 0$.  The functions
$C_{ij}(t)$ are given by the usual weak-coupling expressions
\begin{equation}
C_{ij}(t) = \frac{1}{\pi} \int_0^\infty d\omega J_{ij}(\omega)
	\left\{ \coth\left(\frac{\beta\omega}{2}\right)\cos(\omega t) - i \sin(\omega t) \right\}.
\end{equation}
Observe that in the Markovian limit, Eqs.~(\ref{eq:gammaplus})-(\ref{eq:gammaminus}) reduce to ordinary
one-sided Fourier transforms.

\section{Calculation of spectroscopic observables}\label{app:spec}

Spectroscopic observables, such as absorption and emission lineshapes, can be straightforwardly calculated from
RDM dynamics.  For example, the $\epsilon$-polarized absorption lineshape, $I_{\epsilon}(\omega)$, is known to be given by the
Fourier transform of the dipole-dipole correlation function, $D_{\epsilon}(t)$\cite{may11,ren02}.
Within certain approximations\cite{may11}, the dipole-dipole correlation function can be written as
$D_{\epsilon}(t) = \sum_\alpha |\mu_{\alpha,\epsilon}|^2 \rho_{\alpha 0}(t)$ with $\rho_{\alpha 0}(0) = 1$,
where $\mu_{\alpha,\epsilon}$ is the $\epsilon$ component of the transition dipole moment from the ground
state to the electronic \textit{adiabatic} state $\alpha$, yielding
\begin{equation}\label{eq:abs}
I_\epsilon(\omega) = \sum_{\alpha} |\mu_{\alpha,\epsilon}|^2\ \textrm{Re} \int_0^\infty dt e^{i\omega t} \rho_{\alpha0}(t).
\end{equation}
As a simple example, consider neglecting the
electron-phonon coupling, such that $\rho_{\alpha0}(t) = \exp(-i\omega_{\alpha0}t)$; then the absorption
spectrum takes the familiar form
\begin{equation}\label{eq:abs_el}
\begin{split}
I_\epsilon(\omega) &= \sum_{\alpha} |\mu_{\alpha,\epsilon}|^2\ 
    {\rm Re} \int_0^\infty dt e^{i\omega t} e^{-i\omega_{\alpha0}t} \\
    &= \sum_{\alpha} |\mu_{\alpha,\epsilon}|^2\ \delta(\omega - \omega_{\alpha0}),
\end{split}
\end{equation}
which clearly neglects any broadening or multiphonon effects.  In the general case,
i.e. by propagating the reduced density matrix $\rho(0) = |\alpha\rangle \langle 0|$ under
the Redfield equation, such environmental effects may be included, as will be demonstrated numerically
in Sec.~\ref{sec:redfield}.  Renger and Marcus\cite{ren02} proposed the analytically useful procedure whereby one
keeps the full time dependence of the diagonal coherence dephasing tensor, $\Gamma^{+}_{\alpha\alpha\alpha\alpha}(t)$,
and performs the Markov approximation for the off-diagonal tensors, $\Gamma^{+}_{\alpha\beta\beta\alpha}$,
all within the secular approximation, i.e.
\begin{equation}
\frac{d\rho_{\alpha 0}(t)}{dt} = -i\omega_{\alpha 0}\rho_{\alpha 0}(t)
	+ R_{\alpha 0 \alpha 0}(t) \rho_{\alpha 0}(t),
\end{equation}
with
\begin{equation}
R_{\alpha 0 \alpha 0}(t) \approx -\Gamma^{+}_{\alpha\alpha\alpha\alpha}(t)
    -\sum_{\beta} \Gamma^{+}_{\alpha\beta\beta\alpha}(t\rightarrow \infty).
\end{equation}

Before concluding, we consider a specific example of a spectroscopic calculation within the time-dependent Redfield formalism,
relevant to the absorption of a single molecule, the so-called pure dephasing problem\cite{ski86,rei96}.
In this case, there is no electronic coupling, $V_{ij} = 0$, and the Hamiltonian is completely specified
by the two level system,
\begin{equation}
\begin{split}
\hat{H}_{tot} &= |S_0\rangle E(S_0) \langle S_0| + |S_1\rangle \left( E(S_1) + \sum_k c_k q_k \right) \langle S_1| \\
&\hspace{1em} + \sum_k \left( \frac{p_k^2}{2} + \frac{1}{2}\omega_k^2 q_k^2 \right).
\end{split}
\end{equation}
The time-local Redfield equation of motion for the coherence $\rho_{S_1S_0}(t)$ is simply
\begin{equation}
\begin{split}
\frac{d\rho_{S_1S_0}(t)}{dt} &= - i \left[ E(S_1) - E(S_0) \right] \rho_{S_1 S_0}(t) \\
&\hspace{1em} + R_{S_1S_0S_1S_0}(t) \rho_{S_1 S_0}(t),
\end{split}
\end{equation}
which can be straightforwardly solved. With the initial condition $\rho(0) = |S_1\rangle \langle S_0|$, one finds
\begin{equation}\label{eq:coh}
\rho_{S_1S_0}(t) = \exp\left[-i (\omega_{S_1S_0} - \lambda_{S_1}) t + g(t) - g(0) \right],
\end{equation}
where the lineshape function $g(t)$ is given by
\begin{equation}
g(t) = \frac{1}{\pi} \int_0^\infty d\omega \frac{J_{S_1}(\omega)}{\omega^2}
	\left\{ \coth\left(\frac{\beta\omega}{2}\right)\cos(\omega t) - i \sin(\omega t) \right\}.
\end{equation}

\begin{figure}[t]
\centering
\includegraphics[scale=1.0]{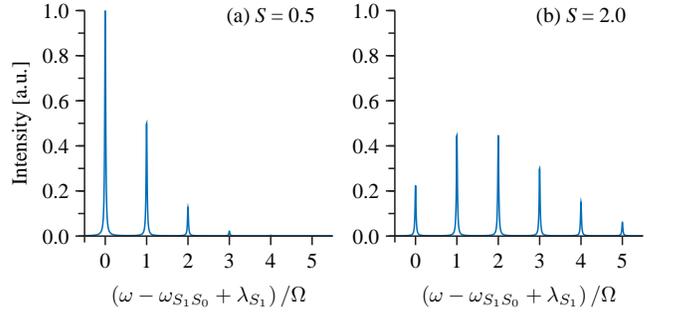}
\caption{
Single molecule absorption spectra at $T=0$ for a single vibrational frequency, $\Omega$, with the
Huang-Rhys parameter, $S$, as given.  Spectra have been artificially broadened for clarity
and normalized to the value of the $S=0.5$
zero-phonon (0-0) line.
}
\label{fig:abs}
\end{figure}

In fact, the pure dephasing problem can be straightforwardly solved exactly\cite{rei96}, using the fact
that the Hamiltonian is already diagonal in the electronic states and well-known thermal
properties of harmonic oscillators.  If one carries out this procedure, it is found to give exactly
the same result as Eq.~(\ref{eq:coh}), a remarkable property of the time-local Redfield equation.
Thus, the cumulant resummation inherent in the time-local formalism is exact for the pure dephasing problem.
This result does not hold for the the time-nonlocal approach.

To make our example more specific, consider a single vibrational mode at frequency $\Omega$,
$J(\omega) = \pi S\Omega^2 \delta(\omega-\Omega)$, such that
\begin{equation}
g(t) = S \left[ \coth(\beta \Omega/2) \cos(\Omega t) - i\sin(\Omega t) \right].
\end{equation}
We have introduced the dimensionless Huang-Rhys factor,
$S = \pi^{-1} \int_0^\infty d\omega J(\omega)/\omega^2$
making clear that $g(0)$ is, at zero temperature, identical to the Huang-Rhys factor.
Although an analytical evaluation of the required Fourier integral, Eq.~(\ref{eq:abs}), is still
difficult, it can be straightforwardly calculated numerically.
We show in Fig.~\ref{fig:abs} example absorption spectra calculated for two different values of the
Huang-Rhys parameter, $S$. Clearly the well-known vibronic progression is perfectly captured, even
in regimes of very strong system-bath coupling.
Again we emphasize that this nonperturbative multi-phonon behavior is a purely non-Markovian effect which is only
captured exactly by the time-local form of the Redfield equation.  The Markovian limit would
yield only a single Lorentzian lineshape at
$\omega = \omega_{S_1S_0} + \textrm{Im} \int_0^\infty ds C_{S_1}(s)$, with broadening
$\textrm{Re} \int_0^\infty ds C_{S_1}(s)$.

\bibliography{fission}

\end{document}